\documentclass[prd,aps,reprint,noeprint,superscriptaddress,amsmath, amssymb]{revtex4-1}
\usepackage{graphicx}% Include figure files
\usepackage{subfigure}
\usepackage{lineno}
\usepackage{epsfig}
\usepackage{multirow}
\usepackage{overpic}
\usepackage{color}
\usepackage{bm} %用于加粗公式中的符号
\usepackage{xspace} %空格相关的
\usepackage{dcolumn}% Align table columns on decimal point
\usepackage{booktabs,tabularx}
\usepackage{array}
\usepackage{textcomp}
\usepackage[colorlinks,linkcolor=blue,urlcolor=blue,citecolor=blue]{hyperref}
\usepackage{tikz}

\usepackage{cleveref}%multiple figure reference

\def\ra{\ensuremath{\rightarrow}}%  "GOES TO" arrow.

\newcommand{\jpsi}{J/\psi}

\newcommand{\piz}{\pi^0}
\newcommand{\Lambdagn}{\Lambda\to n\gamma}
\newcommand{\Lambdapizn}{\Lambda\to n\pi^{0}}

\newcommand{\Lambdapipb}{\bar\Lambda\to \bar{p}\pi^{+}}
\newcommand{\Lambdagnb}{\bar\Lambda\to \bar{n}\gamma}
\newcommand{\Lambdapip}{\Lambda\to {p}\pi^{-}}
\newcommand{\gev}{\rm GeV}
\newcommand{\gevc}{\rm{GeV}/\emph{c}}
\newcommand{\gevcc}{\rm{GeV}/\emph{c}^{2}}
\newcommand{\mevcc}{\rm{MeV}/\emph{c}^{2}}
\newcommand{\jpsill}{\jpsi\to\Lambda\bar\Lambda}
\begin{document}
\normalsize
\parskip=5pt plus 1pt minus 1pt

%%%%%%%%%%%%%%%%%%%%%%%%%%%%%%%%%%%%%%%%%%%%%%%%%%%%%%%%%%%%%%%%%

\title{\boldmath Measurement of the branching fraction and decay asymmetry of $\Lambda\to n\gamma$}
\author{
  \begin{center}
    \begin{small}
M.~Ablikim$^{1}$, M.~N.~Achasov$^{11,b}$, P.~Adlarson$^{70}$, M.~Albrecht$^{4}$, R.~Aliberti$^{31}$, A.~Amoroso$^{69A,69C}$, M.~R.~An$^{35}$, Q.~An$^{66,53}$, X.~H.~Bai$^{61}$, Y.~Bai$^{52}$
% ...
    \\
    \vspace{0.2cm}
    (BESIII Collaboration)\\
    \vspace{0.2cm}
    {\it$^{1}$ Institute of High Energy Physics, Beijing 100049, People's Republic of China\\
% ...
$^{4}$ Bochum Ruhr-University, D-44780 Bochum, Germany\\
% ...
$^{11}$ G.I. Budker Institute of Nuclear Physics SB RAS (BINP), Novosibirsk 630090, Russia\\
% ...
$^{31}$ Johannes Gutenberg University of Mainz, Johann-Joachim-Becher-Weg 45, D-55099 Mainz, Germany\\
% ...
$^{35}$ Liaoning Normal University, Dalian 116029, People's Republic of China\\
% ...
$^{52}$ Southeast University, Nanjing 211100, People's Republic of China\\
$^{53}$ State Key Laboratory of Particle Detection and Electronics, Beijing 100049, Hefei 230026, People's Republic of China\\
% ...
$^{61}$ University of Jinan, Jinan 250022, People's Republic of China\\
% ...
$^{66}$ University of Science and Technology of China, Hefei 230026, People's Republic of China\\
% ...
$^{69}$ University of Turin and INFN, (A)University of Turin, I-10125, Turin, Italy; (B)University of Eastern Piedmont, I-15121, Alessandria, Italy; (C)INFN, I-10125, Turin, Italy\\
$^{70}$ Uppsala University, Box 516, SE-75120 Uppsala, Sweden\\
% ...
$^{b}$ Also at the Novosibirsk State University, Novosibirsk, 630090, Russia\\
% ...
    }
  \vspace{0.4cm}
  \end{small}
  \end{center}
  }

%%%%%%%%%%%%%%%%%%%%%%%%%%%%%%%%%%%%%%%%%%%%%%%%%%%%%%%%%%%%%%%%%
\begin{abstract}
  The radiative hyperon decay $\Lambdagn$ is studied using
  $(10087\pm44)\times 10^6$ $\jpsi$ events collected with the BESIII
  detector operating at BEPCII.  The absolute branching fraction of
  the decay $\Lambdagn$ is determined with a significance of
  5.6$\sigma$ to be $[0.832\pm0.038(\rm stat.)\pm0.054(\rm
  syst.)]\times10^{-3}$, which lies significantly below the current
  PDG value. By analyzing the joint angular distribution of the decay
  products, the first determination of the decay asymmetry
  $\alpha_{\gamma}$ is reported with a value of $-0.16\pm0.10(\rm
  stat.)\pm0.05(\rm syst.)$.

\end{abstract}

%\pacs{13.25.Gv, 12.38.Qk, 14.20.Gk, 14.40.Cs}
\maketitle
%\linenumbers

%%%%%%%%%%%%%%%%%%%%%%%%%%%%%%%%%%%%%%%%%%%%%%%%%%%%%%%%%%%%%%%%%
%\input{Contents}
%%%%%%%%%%%%%%%%%%%%%%%%%%%%%%%%%%%%%%%%%%%%%%%%%%%%%%%%%%%%%%%%%%%%%%%%%%%%%%%
Weak radiative transitions of hadrons are governed by the interplay of
the electromagnetic, weak, and strong interactions~\cite{Lach1995} and
involve parity violating (p.v.) and parity conserving (p.c.)
amplitudes. According to Hara's theorem~\cite{Hara1964}, the
p.v. amplitude of radiative hyperon decays, $B_i\to B_f\gamma$,
vanishes in the limit of SU(3) flavor symmetry. Taking into account
the breaking of this symmetry in the quark model, the decay asymmetry
$\alpha_\gamma$, given by the interference between p.v. and
p.c. amplitudes, is expected to be positive for decays such as
$\Sigma^{+}\to p\gamma$, where the $s$ quark in the initial state
baryon decays to a $d$ quark. It was, therefore, a surprise when
several experiments reported a large negative value of the decay
asymmetry for this
process~\cite{Manz1980,Kobayashi1987777,Foucher1992,Foucher1992akkk,Timm1995}.
For other radiative hyperon decays, measurements have found
non-vanishing positive decay asymmetries~\cite{James1990,Teige1989}.
The disagreement between theoretical expectation and experimental
results provoked wide interest in these processes, and various
solutions to the puzzle were
proposed~\cite{Vasanti1976uuu,Kogan1983sss,Gilman1979asas,Goldman1989,Scadron1983,Palle1987,Zenczykowski1991}.
It was suggested that the validity of Hara's theorem could be
confirmed by determining the sign of the $\Lambda\to n\gamma$ decay
asymmetry~\cite{Zenczykowski2020a}, a positive value indicating the
theorem's violation~\cite{Gavela1981a,Zenczykowski2006a,Niu2021zhao}.

In the three previous measurements of $\Lambda\to n\gamma$ performed
by two fixed target
experiments~\cite{Biagi:1986vn,Noble1992,Larson1993a}, the branching
fraction~(BF) was obtained from the ratio
$\mathcal{B}_{\Lambdagn}/\mathcal{B}_{\Lambdapizn}$. Only the result
from Ref.~\cite{Larson1993a} is considered by the Particle Data Group
(PDG)~\cite{Zyla2020}.  The decay asymmetry of $\Lambda\to n\gamma$,
however, which is essential for the test of the Hara theorem, has not
been measured so far.

At BESIII, a measurement of the $\Lambdagn$ decay utilizing the large yield of $\Lambda\bar\Lambda$ pairs from $\jpsill$~\cite{Li2017haibo} is accomplished using a double-tag~(DT) technique~\cite{Baltrusaitis1986}. The $\jpsill$ events are identified by reconstructing the pionic decay $\Lambdapipb$~($\Lambdapip$),  denoted as single-tag (ST). Then a DT event consisting of an ST $\bar\Lambda$~($\Lambda$) candidate accompanied with a $\Lambdagn$~($\Lambdagnb$) candidate is selected. The absolute BF of the decay $\Lambdagn$ is given by
\begin{equation}
\label{eq:branching ftaction}
     \mathcal{B}_{\Lambdagn} =  \frac{N_{\rm DT}/\varepsilon_{\rm DT}} {N_{\rm ST}/\varepsilon_{\rm ST}}, 
\end{equation}
where $N_{\rm ST}~(N_{\rm DT})$ and $\varepsilon_{\rm ST}~(\varepsilon_{\rm DT})$ are the ST~(DT) yield and the corresponding detection efficiency. Here and throughout this letter, charge-conjugate channels are implied unless stated otherwise.

A previous BESIII study~\cite{BESIIIPolirization2019} showed that the
$\Lambda$ from $\jpsill$ is transversely polarized with a magnitude
reaching 25\%. This polarization can be used to determine the decay
asymmetry $\alpha_\gamma$ in the $\Lambdagn$ decay from the angular
distribution of the daughter baryons from the $\jpsill$
process~\cite{Faeldt2017}. Generally, the joint angular distribution
$\mathcal{W}$ of $J/\psi\to\bar{\Lambda}(\to
\bar{p}\pi^{+})\Lambda(\to n\gamma)$ can be expressed as:
\begin{small}
\begin{equation}
\label{joint_angular_eq}
\begin{aligned}
&\mathcal{W}\left(\xi ; \alpha_{\psi}, \Delta \Phi, \alpha_{\gamma}, \alpha_{+}\right) \\
&=1+\alpha_{\psi} \cos ^{2} \theta_{\Lambda}+\alpha_{\gamma} \alpha_{+}\left[\sin ^{2} \theta_{\Lambda}\left(n_{1}^{x} n_{2}^{x}-\alpha_{\psi} n_{1}^{y} n_{2}^{y}\right)\right. \\
&\left.+\left(\cos ^{2} \theta_{\Lambda}+\alpha_{\psi}\right) n_{1}^{z} n_{2}^{z}\right] \\
&+\alpha_{\gamma} \alpha_{+} \sqrt{1-\alpha_{\psi}^{2}} \cos (\Delta \Phi) \sin \theta_{\Lambda} \cos \theta_{\Lambda}\left(n_{1}^{x} n_{2}^{z}+n_{1}^{z} n_{2}^{x}\right) \\
&+\sqrt{1-\alpha_{\psi}^{2}} \sin (\Delta \Phi) \sin \theta_{\Lambda} \cos \theta_{\Lambda}\left(\alpha_{\gamma} n_{1}^{y}+\alpha_{+} n_{2}^{y}\right),
\end{aligned}
\end{equation}
\end{small}where $\hat{\textbf{n}}_{1}$~( $\hat{\textbf{n}}_{2}$) is the unit vector in the direction of the neutron (anti-proton) in the $\Lambda$~($\bar\Lambda$) rest frame. The components of $\hat{\textbf{n}}_{1}$ and $\hat{\textbf{n}}_{2}$ are $(n^{x}_{1},n^{y}_{1},n^{z}_{1})$ and $(n^{x}_{2},n^{y}_{2},n^{z}_{2})$, in a coordinate system where the $z$ axis of both the $\Lambda$ and the $\bar\Lambda$ rest frame is oriented along the momentum $\mathbf{p}_{\Lambda}$ at an angle $\theta_{\Lambda}$ with respect to the $e^-$ beam direction. The $y$ axis is perpendicular to the production plane and oriented along the vector $\mathbf{k}_{-} \times \mathbf{p}_{\Lambda}$, where $\mathbf{k}_{-}$ is the $e^{-}$ beam momentum in the $\jpsi$ rest frame. For each event, the full set of kinematic variables~($\theta_{\Lambda}, \hat{\textbf{n}}_{1}, \hat{\textbf{n}}_{2}$) is denoted by $\xi$. Furthermore, $\alpha_{\psi}$ and $\Delta \Phi$ denote the absolute ratio of the two helicity amplitudes of $\jpsill$ and their relative phase, respectively, and $\alpha_{\gamma}$ ($\alpha_{+}$) is the decay asymmetry of $\Lambdagn$ ($\Lambdapipb$ ). 

In this letter, we report the absolute BF and the decay asymmetry of
$\Lambda\to n\gamma$ from $(10087\pm44)\times 10^6$ $\jpsi$
events~\cite{jpsinumber2021} collected at the BESIII
detector~\cite{Ablikim2010,Ablikim2020lalal} operating at the BEPCII
collider~\cite{Yu2016}. A detailed description of the BESIII detector
can be found in Ref.~\cite{Ablikim2010}.  Simulated data samples
produced with {\sc Geant4}-based~\cite{G42002iii} Monte Carlo~(MC)
software, including a detailed geometric description of the BESIII
detector and the detector response, are used to determine detection
efficiencies and estimate background contributions. The $\jpsi$
resonance is generated by {\sc kkmc}~\cite{Jadach2001}, incorporating
the effects of the beam energy spread and the initial state radiation
in the $e^+e^-$ annihilation. The subsequent decays are modeled with
{\sc EvtGen}~\cite{Lange2001,*Ping2008} using the BFs taken from the
PDG~\cite{Zyla2020} for known decays and {\sc
  LundCharm}~\cite{Chen2000,*Yang2014a} for remaining unknown decays.
A sample of simulated $\jpsi$ decay events~(the inclusive MC sample),
corresponding to the luminosity of data, is used to study background
events. Signal MC samples, including a sample of $5.6\times10^{7}$
$\jpsi\to\bar\Lambda~(\to \bar{p}\pi^{+})\Lambda~(\to\rm anything)$
and a sample of $4\times10^{5}$ $\jpsi\to\bar\Lambda~(\to
\bar{p}\pi^{+})\Lambda~(\to n\gamma)$, are generated to estimate the
ST and DT signal efficiencies, respectively. The joint angular
distributions are generated according to Eq.~\eqref{joint_angular_eq},
where $\alpha_{\gamma}$ is adopted from this analysis and
$\alpha_{\psi}= 0.461\pm0.006\pm0.007$, $\Delta \Phi =
42.4\pm0.6\pm0.5^{\circ}$ and $\alpha_{+} = -0.758\pm0.010\pm0.007$
from Ref.~\cite{BESIIIPolirization2019}.  Moreover, a sample of
$2\times10^{7}$ $\jpsi\to\bar\Lambda~(\to \bar{p}\pi^{+})\Lambda~(\to
n\pi^0)$ events is generated to study the dominant background.

The ST $\bar\Lambda$ candidate is reconstructed through the dominant
decay mode $\Lambdapipb$. Charged tracks are detected in the main
drift chamber~(MDC) and must satisfy the condition
$|\cos\theta|<0.93$, where $\theta$ is the polar angle with respect to
the MDC symmetry axis.  The momenta ranges of pions and anti-protons
from the $\bar\Lambda$ decays are well separated, thus the tracks with
momenta less than 0.5~$\gevc$ are assigned to be pions, otherwise
anti-protons. In addition, measurements of the specific ionization
energy loss in the MDC and the flight time by the time-of-flight
system are combined to perform particle identification for the
(anti-)proton candidate. They are required to have the largest
likelihood for the particle type selected among the pion, kaon and
proton hypotheses.  Events with at least one anti-proton and one
positively charged pion are selected.  A vertex fit is performed to
each $\bar{p}\pi^+$ pair, and the one with the minimum $\chi_{\rm
  vtx}^{2}$ of the vertex fit is regarded as the $\bar{\Lambda}$
candidate for further analysis. The $\bar{\Lambda}$ candidate is
required to have $\chi_{\rm vtx}^{2}$ less than 20, an invariant mass
within 8~$\mevcc$ of the nominal $\Lambda$ mass~\cite{Zyla2020} and
a decay length relative to the interaction point larger than twice
of its resolution.

\begin{figure}[htbp]
\subfigure
{
  \begin{overpic}
    [scale=0.35]{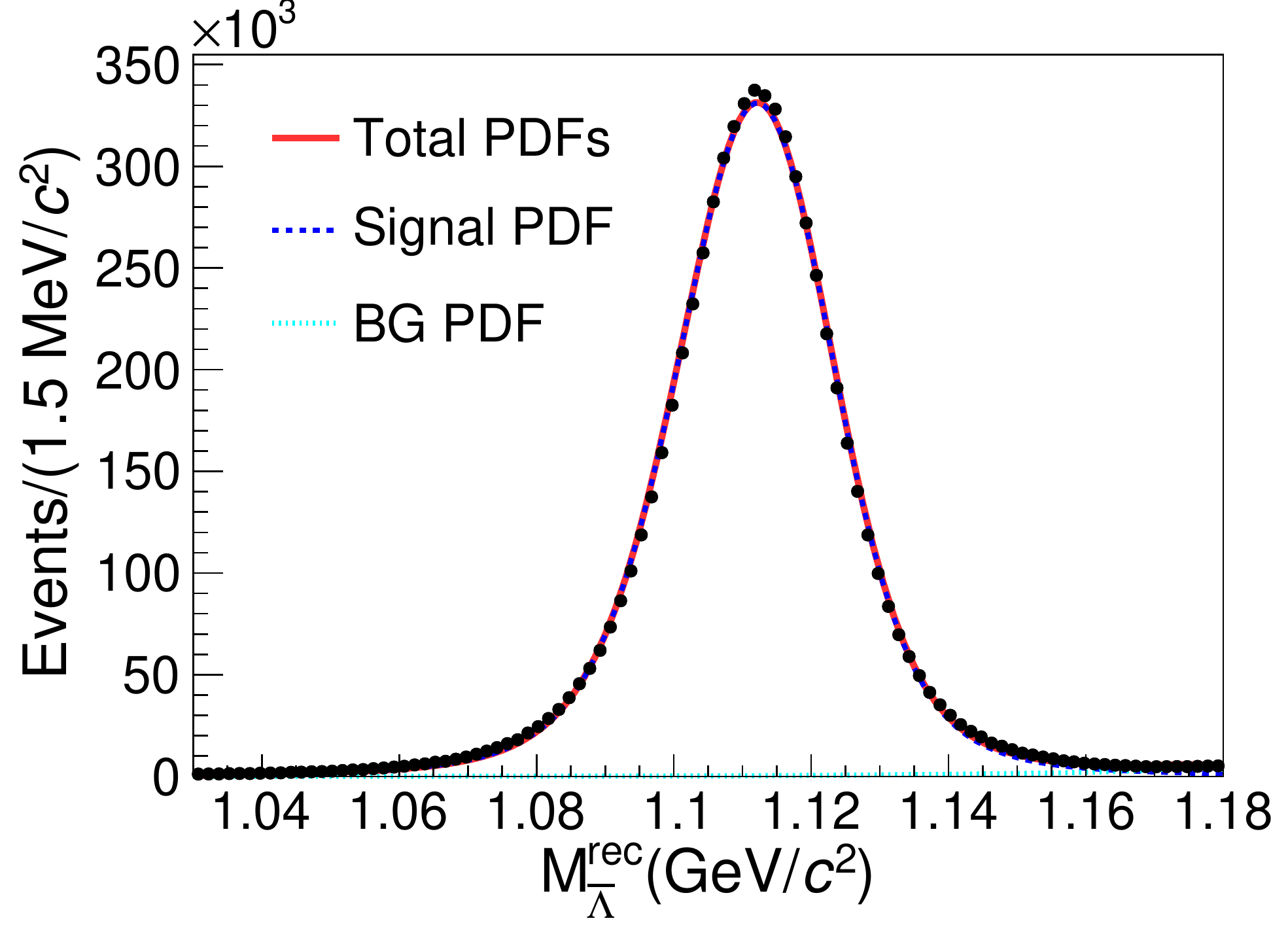}\put(75,60){}
  \end{overpic}
}
	\caption{Distribution of the recoil mass $M^{\rm rec}_{\bar\Lambda}$ after ST
	selection. The black dots with error bars represent data, the error bars are smaller than the marker size. The red solid line shows the fit result and the blue dashed and cyan dotted lines are the signal and background distributions, respectively.
}
\label{fig_ST_fit}
\end{figure}

To identify events with $\jpsill$ and reduce the background
contributions from $\jpsi\to \bar{\Lambda}+\rm anything$ which are not
due to $\jpsill$, a recoil mass $M^{\rm rec}_{\bar\Lambda} = \sqrt{
  (E_{\rm c.m.} - E_{\bar\Lambda})^{2} - P_{\bar\Lambda}^{2}}$ is
defined, where $E_{\rm c.m.}$ is the center-of-mass~(c.m.) energy,
$E_{\bar\Lambda}$ is the energy and $P_{\bar\Lambda}$ the momentum of
the ST $\bar\Lambda$ candidate. This mass is required to be within
$1.03<M^{\rm rec}_{\bar\Lambda} <1.18$~$\gevcc$. The distribution of
$M^{\rm rec}_{\bar\Lambda}$ is shown in Fig.~\ref{fig_ST_fit}, where
only few background events are observed.  A maximum likelihood fit is
performed to determine the signal yield, where the signal and background
distributions are represented by shapes obtained from signal MC and
inclusive MC samples. The MC shapes are convolved with a Gaussian
function to account for imperfect simulation of the detector
resolution. The charge conjugate channels are analyzed with the
duplicate processing method, and the yields of ST $\Lambda$ and
$\bar{\Lambda}$ candidates from the fits are summarized in
Table~\ref{tab:fit_results}. The background contribution is less than
1\%, which is also validated by the inclusive MC sample.

%%%fah

On the signal side we search for $\Lambdagn$ from the residual neutral particles in the ST $\bar{\Lambda}$ candidates. 
%The signal $\Lambdagn$ is reconstructed with the residual neutral particles in the ST $\bar{\Lambda}$ candidates. %Events are required to contain one $\bar{p}$ and one $\pi^+$.
For a neutral shower, the deposited energy in the electromagnetic calorimeter~(EMC) should be larger than 25~MeV in the barrel region ($|\cos\theta| < 0.80$) or 50~MeV in the end cap region ($0.86<|\cos\theta|<0.92$).
To reject secondary showers originating from charged tracks, the
shower candidates are required to be apart from the charged tracks
with an opening angle of $10^\circ$ for pion and proton tracks and
$20^\circ$ for anti-proton tracks.  To suppress electronic noise, the
interval between the EMC response time and the event start time is
required to be within 700~ns.  There are two neutral particles in the
final states of the signal process $\Lambdagn~(\Lambdagnb)$, the
photon and the (anti-)neutron. The radiative photon produces a shower
in the EMC with deposited energy less than 400~MeV. With a probability
of 0.65, the $\bar n$ annihilates in the EMC and produces several
secondary particles. The most energetic shower with energy deposit
larger than 0.4~GeV is regarded as an $\bar n$ candidate. The $n$,
meanwhile, is difficult to identify due to its low interaction
efficiency and its small energy deposition, and is treated as a
missing particle. Therefore, only the $\gamma$s and $\bar n$ are
selected in this analysis.  At least one shower is required as a
$\gamma$ candidate in an event for $\Lambdagn$, and at least two
showers as $\gamma$ and $\bar n$ in an event for $\Lambdagnb$. For the
reconstruction of $\Lambdagn$, a one-constraint~(1C) kinematic fit is
applied by imposing energy-momentum conservation of the candidate
particles in the hypothetical $\jpsi\to\bar\Lambda n\gamma$ process,
where the neutron is set as a missing particle. On the other hand, for
the reconstruction of $\Lambdagnb$, a 3C kinematic fit is imposed for
the $\jpsi\to\Lambda \bar n\gamma$ process, where the direction of the
$\bar{n}$ is measured and the energy is unmeasured. For events with
multiple photon candidates, the combination giving the minimum
$\chi_{\rm 1C}^2$ ($\chi_{\rm 3C}^2$) is retained for the
analysis. Furthermore, $\chi_{\rm 1C}^2$ ($\chi_{\rm 3C}^2$) is
required to be less than 10~(15).

Detailed MC studies show that the dominant background contribution
comes from the $\Lambdapizn$ decay with its large BF, while other
background processes are almost negligible.  The background can be
classified into two categories: first, events with the detected photon
from the $\pi^0$ decay in $\Lambdapizn$, denoted as BG~A, and second,
events with the detected photon not from the $\pi^0$ decay, denoted as
BG~B. In the latter case, the photons arise from noise or a shower
from secondary products of other
particles. 
In order to suppress BG~A, a 1C~(3C) kinematic fit under the
hypothesis $\jpsi\to\bar\Lambda n\gamma\gamma$~($\jpsi\to\Lambda
\bar{n}\gamma\gamma$) is performed, and events surviving the kinematic
fit and with a $\gamma\gamma$ invariant mass within 20~$\mevcc$ of
the $\piz$ nominal mass~\cite{Zyla2020} are
rejected. 
To suppress BG~B, the detected photon is required to have an
energy larger than 150~MeV and an opening angle larger than 20$^\circ$
from the (anti-)neutron candidate.  Additionally, for BG~A and BG~B a
boosted decision tree~(BDT) is applied on the detected photon to
discriminate signal photons from other showers, based on the measured
variables, $i.e.$ deposited energy, secondary moment, number of hits,
Zernike moment~(A$_{42}$), and deposition
shape~\cite{Zernikemoment}. The response of the BDT output is required
to be larger than 0.3, after which 86.8$\%$~(92.8$\%$) of the BG~A and
99.5$\%$~(99.7$\%$) of the BG~B events are rejected with
44.6$\%$~(48.4$\%$) loss of the signal efficiency for the
$\Lambdagn~(\Lambdagnb)$ process.

\begin{figure}[htbp]
	\begin{center}
\begin{tikzpicture}[scale=1.0]
 \node(a) at (-1.0,0.0)
  {\includegraphics[width=1.0\linewidth]{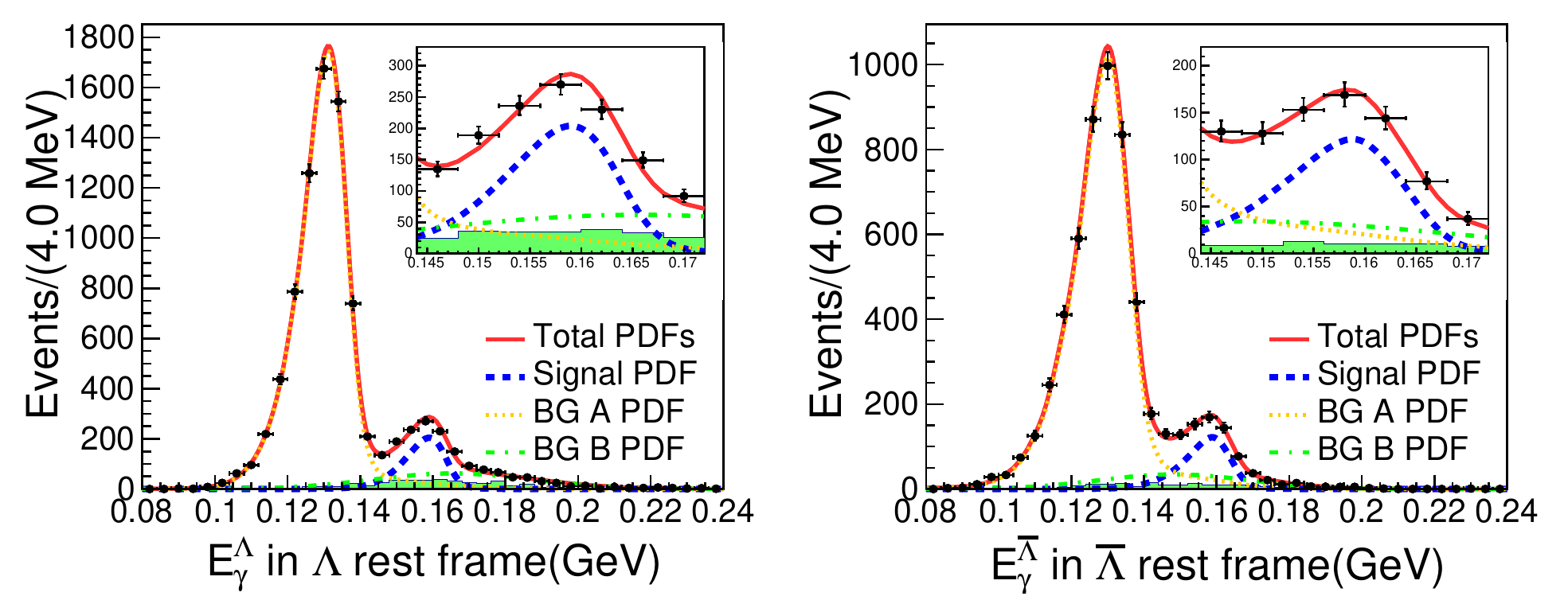}\put(-220,72){\textbf{(a)}}\put(-96,72){\textbf{(b)}}}; 
\end{tikzpicture}  
	\caption{Distributions of $E_{\gamma}^{\Lambda}$ for (a)~$\Lambdagn$ and (b)~$\Lambdagnb$ in the $\Lambda$ and $\bar\Lambda$ rest frame, respectively. The black dots with error bars represent data. The red solid, blue dashed, orange dotted, and green dash-dotted lines denote the fit result, signal, BG~A, and BG~B contributions, respectively. The green histograms indicate the BG~B from an inclusive MC sample. The insets show the details of the fit in the signal region.  
}
\label{fig_DT_fit}
	\end{center}
	\end{figure}

The distribution of the photon energy in the $\Lambda$ rest frame
$E_{\gamma}^{\Lambda}$ after all selection criteria is shown in
Fig.~\ref{fig_DT_fit} for the decay $\Lambdagn$~($\Lambdagnb$), where
the predominant peak around 0.13~$\gev$ is from BG~A, and the
second peak around 0.15~$\gev$ corresponds to the signal.  To
determine the DT signal yields, an unbinned extended maximum
likelihood fit is performed to the $E_{\gamma}^{\Lambda}$
distributions. The signal and BG~A are modeled by the MC simulated
shape convolved with a Gaussian function.  Since BG~B involves a fake
photon and is difficult to be modeled by the MC simulation, its
lineshape is obtained by a data-driven approach with a control sample
of $\Lambdapizn(\to \gamma\gamma)$, and the photon candidates used in
the kinematic fit are from noise photons in the EMC rather than the
two signal photons from $\pi^0\to \gamma\gamma$.
The DT yields obtained from fits are summarized in Table~\ref{tab:fit_results}. The BFs determined according to Eq.~\eqref{eq:branching ftaction} are found to be consistent for the two charge-conjugate modes. Therefore, a simultaneous fit, assuming the same BF for the two modes, is performed, and the results are given in bold font in the Table~\ref{tab:fit_results}.

\begin{table}[htbp]
\renewcommand\arraystretch{1.2}
\caption{The results for the decays $\Lambdagn$ and $\Lambdagnb$. The BF and $\alpha_{\gamma}$ values are given both for individual and simultaneous fits. The first (second) uncertainties are statistical (systematic).}
\label{tab:fit_results}
\centering
\scalebox{1.0}
{
\begin{tabular}{l c c}
\hline \hline 
\noalign{\vskip 3pt}
Decay Mode   & ~~~~~$\Lambdagn$~~~~~~ & ~~~~~$\Lambdagnb$~~~~\\
\hline
$N_{\rm ST}$ ~($\times10^{3}$)&   $6853.2\pm2.6$    &   $7036.2\pm2.7$  \\
$\varepsilon_{\rm ST}~(\%)$&  51.13$\pm$0.01  & 52.53$\pm$0.01  \\
$N_{\rm DT}$   & 723$\pm$40  & 498$\pm$41  \\
$\varepsilon_{\rm DT}~(\%)$  &  6.58$\pm$0.04  & 4.32$\pm$0.03 \\ 
\hline
\multirow{2}{*}{BF~($\times10^{-3}$)}  &~~~0.820$\pm$0.045$\pm$0.066~~~ & 0.862$\pm$0.071$\pm$0.084~~~ \\
     &  \multicolumn{2}{c}{\textbf{0.832$\pm$0.038$\pm$0.054}} \\
\hline
\multirow{2}{*}{$\alpha_{\gamma}$} & $-0.13\pm$0.13$\pm$0.03  &   0.21$\pm$0.15$\pm$0.06  \\
     &  \multicolumn{2}{c}{\textbf{$-$0.16$\pm$0.10$\pm$0.05}} \\
\hline
\hline
\end{tabular}
}
\end{table}

The systematic uncertainties on the BF measurement stem from the
photon and anti-neutron detection efficiency, kinematic fit, invariant
$M_{\gamma\gamma}$ mass selection window, opening angle between photon
and (anti-)neutron, BDT output for the photon, MC model due to
$\alpha_{\gamma}$, and fit procedure.  The uncertainties associated
with ST selection almost cancel based on Eq.~\eqref{eq:branching
  ftaction}. There is only one photon for the signal process, and the
uncertainty associated with the photon detection efficiency is $1\%$
according to Ref.~\cite{Ablikimphoton2011}.  The uncertainty of the
anti-neutron detection efficiency is negligible after correcting the
efficiency by a data-driven method~\cite{Liu2022n}.  
To estimate the uncertainty due to the kinematic fit, we change the value of the $\chi^2_{\rm 1C}~(\chi^2_{\rm 3C})$ by $\pm 1$ and investigate the fluctuation on the BF, which is taken as a systematic uncertainty.
%The uncertainty from the kinematic fit is estimated at $2.0\%$ ($1.5\%$) by varying the value of the $\chi^2_{\rm 1C}~(\chi^2_{\rm 3C})$ requirement by 1.0, and the taking the largest change of the BF with respect to the nominal result.
The uncertainty from the
$M(\gamma\gamma)$ mass window requirement is studied with the control
sample of $\jpsi\to\bar\Lambda~(\to \bar{p}\pi^{+})\Lambda~(\to
n\pi^0)\piz~(\to\gamma\gamma)$, and the difference of the efficiency
between the data and MC simulation, $1.2\%$, is taken as the
uncertainty. The uncertainty from the requirement of the opening angle
between photon and (anti-)neutron is estimated to be $2.0\%$ by
varying the selection criteria by two degrees. The uncertainty
associated with the BDT for the photon is negligible after correcting
the efficiency using the control sample of
$\jpsi\to\rho\pi~(\to\pi^{+}\pi^{-}\gamma\gamma)$. The uncertainty of
the MC model due to the input decay asymmetry $\alpha_{\gamma}$ is
estimated to be $0.6\%$ by varying the input value of
$\alpha_{\gamma}$ by its uncertainty.  The systematic
uncertainties from the fit of the $E_{\gamma}^{\Lambda}$ distribution
include those associated with the fit range and the modeling of the
signal and background shapes.  An alternative fit of the
$E_{\gamma}^{\Lambda}$ distribution in the $(0.09, 0.23)$~GeV range is
performed, and the resultant difference in signal yield, 0.3\%, is
taken as the uncertainty. The uncertainties associated with the shape
modeling of the signal and BG~A, are estimated by varying the width of
the Gaussian resolution function within the uncertainties. The
resulting differences of the yields, $0.4\%$ for the signal and
$1.0\%$ for BG~A, are assigned as the systematic uncertainties.  To
estimate the shape modeling uncertainty of BG~B, ensembles of
pseudo-data are generated according to modified BG~B distributions as
allowed by the uncertainties, and the resulting standard deviation of
the signal yields, 4.8\%, is taken as the systematic uncertainty.  The
systematic uncertainty associated with the extraction of the ST yield
is estimated by varying the width of the Gaussian resolution function
within its uncertainties for signal shape modeling and replacing the
inclusive MC shape with a second order polynomial function. The
resulting difference in the ST yield of 2.3\% is taken as the
systematic uncertainty.  By adding all these values in quadrature, the
total systematic uncertainty is estimated to be $6.5\%$.

\begin{figure}[htbp]
	\begin{center}
\begin{tikzpicture}[scale=1.0]
 \node(a) at (-1.0,1.0)
  {\includegraphics[width=1.0\linewidth]{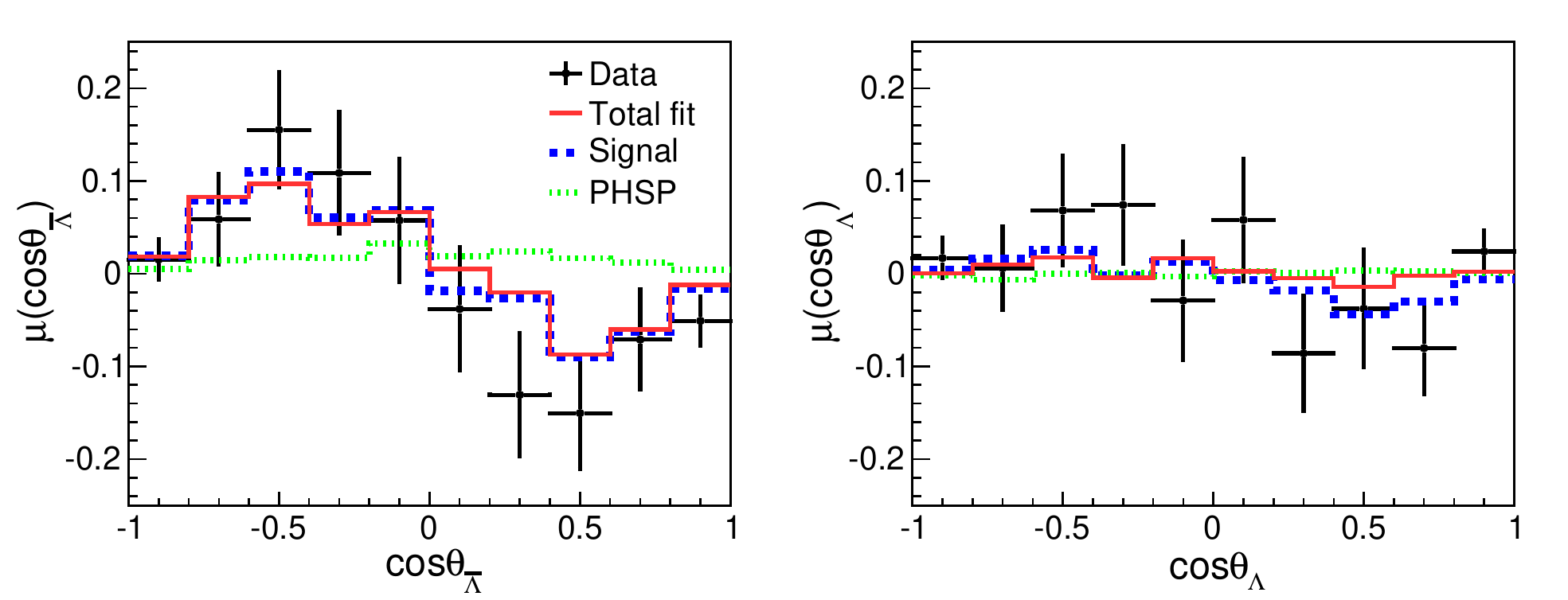}\put(-222,72){\textbf{(a)}}\put(-98,72){\textbf{(b)}}}; 
 \node(b) at (-1.0,-2.3)
  {\includegraphics[width=1.0\linewidth]{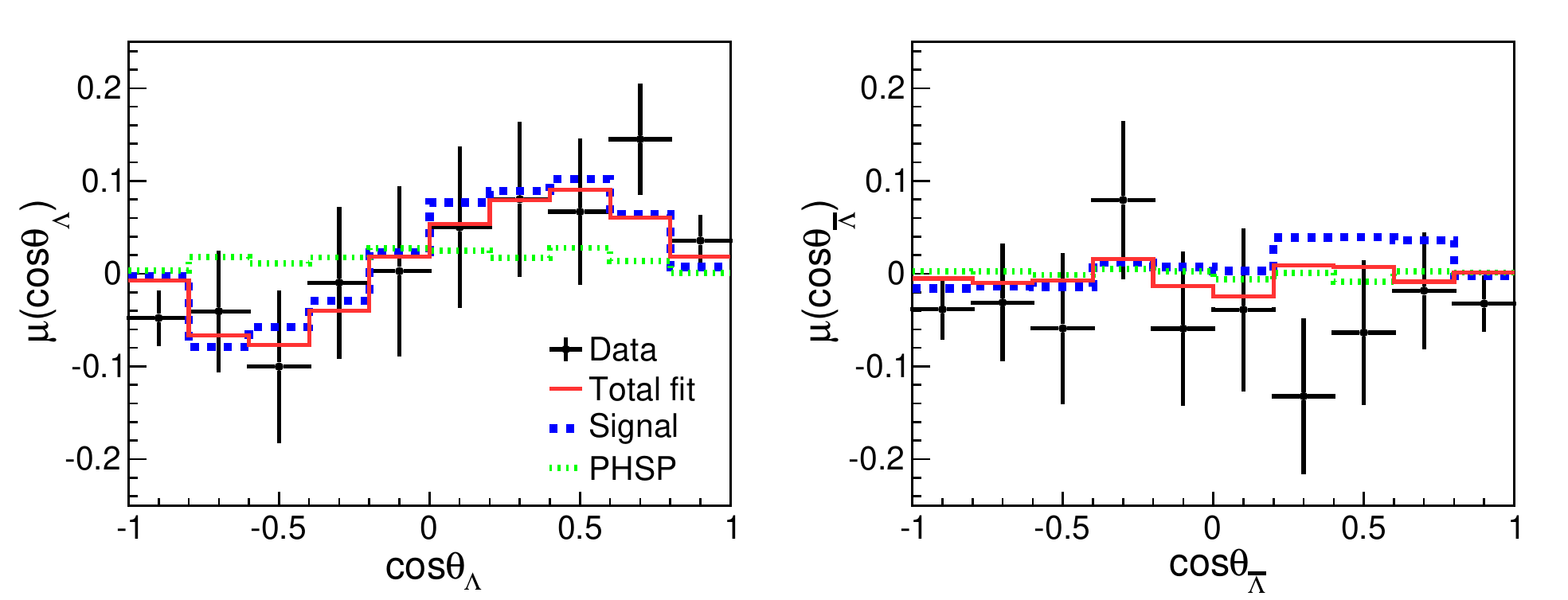}\put(-222,72){\textbf{(c)}}\put(-98,72){\textbf{(d)}}}; 
\end{tikzpicture}  
	\caption{Polarization moment $\mu(\cos\theta_{\bar\Lambda(\Lambda)}$) vs $\cos\theta_{\bar\Lambda(\Lambda)}$ for (a) $\Lambdapipb$, (b) $\Lambdagn$ in the process $\jpsi\to\bar\Lambda~(\to \bar{p}\pi^{+})\Lambda~(\to n\gamma)$, and moment distribution $\mu(\cos\theta_{\Lambda(\bar\Lambda)}$) vs $\cos\theta_{\Lambda(\bar\Lambda)}$ for (c) $\Lambdapip$ and (d) $\Lambdagnb$ in the process $\jpsi\to\Lambda~(\to p\pi^{-})\bar\Lambda~(\to \bar n\gamma)$. %For each event, the moment is calculated using Eq.~(\ref{eql:moment}) for $m=10$ bins in cos$\theta_{\Lambda}$. 
	Dots with error bars indicate data and red solid lines show the fit result. The blue dashed and green dotted lines represent the moment  for signal and PHSP MC, respectively. }
\label{fig_moment_fit}
\end{center}
\end{figure}

The decay asymmetry $\alpha_{\gamma}$ is determined using
Eq.~\eqref{joint_angular_eq} with a maximum likelihood fit. A total of
1994 candidate events within a range of $(0.145, 0.17)$~$\gev$ around
$E_{\gamma}^{\Lambda}$ are used in the fit, with an estimated fraction
of background events of 43.3\%. In the fit of $\alpha_{\gamma}$, the
likelihood function of the $i$-{th} event is calculated through the
probability density function~(PDF):
\begin{equation}
\label{eql:pdf}
\begin{split} 
	 \mathcal{P}(\xi^{i}; \alpha_{\psi}, \Delta \Phi, \alpha_{\gamma}, \alpha_{+}) &= \mathcal{C}\mathcal{W}(\xi^{i}; \alpha_{\psi}, \Delta \Phi, \alpha_{\gamma}, \alpha_{+})\epsilon(\xi^{i}) , 
\end{split} 
\end{equation}
where $\mathcal{C}^{-1}= \int \mathcal{W}(\xi; \alpha_{\psi}, \Delta
\Phi, \alpha_{\gamma}, \alpha_{+}) \epsilon(\xi)d\xi$ is the
normalization factor evaluated by a phase space~(PHSP) MC sample, and
$\alpha_{\psi}, \Delta \Phi, \alpha_{+}$ are fixed to the values in
Ref.~\cite{BESIIIPolirization2019}. The BG~A and BG~B contributions to
the likelihood value are estimated with MC samples and subtracted in
the calculation of the likelihood function. We fit the $\Lambdagn$ and
$\Lambdagnb$ decay modes individually, and the results agree within
statistical uncertainties as summarized in
Table~\ref{tab:fit_results}. A simultaneous fit, assuming the same
magnitude of $\alpha_{\gamma}$ but with opposite sign for the
charge-conjugate modes, is used to determine the decay asymmetry,
yielding $\alpha_{\gamma}(\Lambdagn)=-0.16\pm0.10$, where the uncertainty is
statistical. The polarization is strongly dependent
on the $\Lambda$ direction $\cos\theta_{\Lambda}$ and indicates the
amplitude of the decay asymmetry.  The $n^{y}_{1}$~($n^{y}_{2}$)
moment
\begin{equation}
\label{eql:moment}
\begin{split} 
	 \mu(\cos\theta_{\Lambda}) & =  \frac{m}{N}\sum_{i=1}^{N_k}n^{y}_{1(2)}, \\
\end{split} 
\end{equation} is proportional to the product of the $\Lambda$ polarization and its decay asymmetry. It is calculated for $m=10$ bins in $\cos\theta_{\Lambda}$. Here, $N$ is the total number of events in the data sample and $N_k$ is the number of events in the $k$-{th} $\cos\theta_{\Lambda}$ bin. Figure~\ref{fig_moment_fit} shows the projection of the global fit together with data and PHSP MC results. The fit result for $\Lambdapipb$ clearly deviates from the PHSP curve while the one for $\Lambdagn$ is consistent with PHSP. The difference in magnitude of the moments for $\Lambdapipb$ and $\Lambdagn$ implies different values of the decay asymmetries since the polarization is the same for $\bar\Lambda$ and $\Lambda$.   

The systematic uncertainties on $\alpha_{\gamma}$ are calculated as in
the BF measurement except for the detection of photon and
anti-neutron, as well as the BDT requirements, which are expected to
be negligible.  All the individual uncertainties are estimated by
alternative unbinned maximum likelihood fits with the different
scenarios, and the resultant (maximum) change on $\alpha_{\gamma}$ is
assigned as the corresponding uncertainty.  The uncertainty from the
1C~(3C) kinematic fit is $0.024$~($0.022$), which is estimated by
varying the selection criteria of $\chi^2_{\rm 1C}~(\chi^2_{\rm 3C})$.
The uncertainty associated with the requirement on the invariant
$M_{\gamma\gamma}$ mass window is 0.016 estimated by varying the mass
window from 20 to 27~$\mevcc$.  The uncertainty due to the requirement
of the opening angle between photon and (anti-)neutron is 0.028
estimated by varying the corresponding selection criteria.  The
uncertainty associated with the signal and background magnitudes,
which are determined by the fit of the $E_{\gamma}^{\Lambda}$
distribution, is estimated with the same approaches as in the BF
measurement by changing the fit range, the signal modeling, BG~A and
BG~B modeling, and the resultant differences with respect to the
nominal $\alpha_{\gamma}$ value. The obtained uncertainties are 0.001,
0.001, 0.002, and 0.008, respectively. The uncertainty due to the
fixed parameters~($\alpha_{\psi}, \Delta \Phi, \alpha_{-},\alpha_{+}$)
in the fit is estimated by varying each parameter by
its resolution, and the resulting differences 0.004, 0.09, 0.00, 0.03
are assigned as the uncertainties.
%The uncertainties due to fit range, the signal modeling, BG~A modeling and BG~B modeling are estimated using the same methods as the BF measurement, the resultant differences with respect to the $\alpha_{\gamma}$ nominal value, 0.001, 0.001, 0.002 and 0.008, respectively, are taken as the uncertainties. 
The total systematic uncertainty of $\alpha_{\gamma}$ is estimated to be 0.05 by adding all these uncertainties in quadrature. 

% Two independent $CP$ violating observables are calculated according to Ref.~\cite{Donoghue1986}: 
% \begin{equation}
% \label{eql:cpodd_br}
% \begin{aligned} 
% \Delta_{CP}&= \frac{\mathcal{B}-\mathcal{\bar B}}{\mathcal{B}+\mathcal{\bar B}} = -0.025\pm0.049(\rm stat.)\pm0.063(\rm syst.),\\
% A_{CP} &= \frac{\mathcal{\alpha_{\gamma}}+\mathcal{\bar\alpha_{\gamma}}}{\mathcal{\alpha_{\gamma}}-\mathcal{\bar\alpha_{\gamma}}} = -0.25\pm0.61(\rm stat.)\pm0.16(\rm syst.),
% \end{aligned}
% \end{equation}where the $\mathcal{B}~(\mathcal{\bar B})$ and $\alpha_{\gamma}~(\bar\alpha_{\gamma})$ are the individual results of BF and decay asymmetry from the charge-conjugate channels.

 %\begin{equation}
%\label{eql:cpodd_br}
%\begin{split} 
%	 \Delta_{CP} = \frac{\mathcal{B}(\Lambdagn)-\mathcal{B}(\Lambdagnb)}{\mathcal{B}(\Lambdagn)+\mathcal{B}(\Lambdagnb)} \\
%\end{split} 
%\end{equation}

%\begin{equation}
%\label{eql:cpodd_alpha}
%\begin{split} 
%	  A_{CP} = \frac{\mathcal{\alpha}(\Lambdagn)+\mathcal{\alpha}(\Lambdagnb)}{\mathcal{\alpha}(\Lambdagn)-\mathcal{\alpha}(\Lambdagnb)} \\
%\end{split} 
%\end{equation}

In summary, 
based on the double-tag method, we report the first measurement of the absolute BF of $\Lambdagn$ of $[0.832\pm0.038(\rm stat.)\pm0.054(\rm syst.)]\times10^{-3}$. The measured value of the BF is a factor of two smaller than the previous measurement of~$(1.75\pm0.15)\times10^{-3}$~\cite{Larson1993a}. By analyzing the joint angular distribution of the decay products, the decay asymmetry $\alpha_{\gamma}$ is determined for the first time, at a value of $-0.16\pm0.10(\rm stat.)\pm0.05(\rm syst.)$.

This analysis is the first measurement of radiative hyperon decays at
an electron--positron collider experiment, making use of the huge
number of polarized hyperons produced in $\jpsi$ decays with clean
background. % The presented measurements are critical to our
%understanding of the hyperon semi-leptonic, non-leptonic, and weak
%radiative decays. 
The results for the asymmetry do not agree well
with predictions such as the broken SU(3) pole
model~\cite{Zenczykowski2006a}, Chiral Perturbation
Theory~\cite{Borasoy1999aaa}, or the non-relativistic constituent quark
model~\cite{Niu2021zhao}. Our BF value is consistent with
the lower unitary bound obtained by considering contributions of
$\Lambda\to p\pi^-$ and $\Lambda\to n\pi^0$ weak hadronic decays
together with $p\pi^-\to n\gamma$ and $n\pi^0\to n\gamma$
rescattering, respectively~\cite{Farrar1971}.
%Together the results provide an essential input to constrain the p.v. amplitudes~\cite{Vasanti1976uuu,Kogan1983sss,Gilman1979asas}.

The BESIII collaboration thanks the staff of BEPCII and the IHEP computing center for their strong support. This work is supported in part by National Key R$\&$D Program of China under Contracts Nos. 2020YFA0406400, 2020YFA0406300; National Natural Science Foundation of China~(NSFC) under Contracts Nos. 11635010, 11735014, 11835012, 11935015, 11935016, 11935018, 11961141012, 12022510, 12025502, 12035009, 12035013, 12192260, 12192261, 12192262, 12192263, 12192264, 12192265; the Chinese Academy of Sciences~(CAS) Large-Scale Scientific Facility Program; Joint Large-Scale Scientific Facility Funds of the NSFC and CAS under Contract No. U1832207; CAS Key Research Program of Frontier Sciences under Contract No. QYZDJ-SSW-SLH040; 100 Talents Program of CAS; INPAC and Shanghai Key Laboratory for Particle Physics and Cosmology; ERC under Contract No. 758462; European Union's Horizon 2020 research and innovation programme under Marie Sklodowska-Curie grant agreement under Contract No. 894790; German Research Foundation DFG under Contracts Nos. 443159800, Collaborative Research Center CRC 1044, GRK 2149; Istituto Nazionale di Fisica Nucleare, Italy; Ministry of Development of Turkey under Contract No. DPT2006K-120470; National Science and Technology fund; STFC~(United Kingdom); The Royal Society, UK under Contracts Nos. DH140054, DH160214; Polish National Science Centre through the Grant No. 2019/35/O/ST2/02907; The Swedish Research Council; U. S. Department of Energy under Contract No. DE-FG02-05ER41374. National Key R$\&$D Program of China under Contracts Nos. 2020YFA0406400, 2020YFA0406300; National Natural Science Foundation of China~(NSFC) under Contracts Nos. 11335008,11625523, 12035013, 11705192, 11950410506, 12061131003, 12105276, 12122509; Joint Large-Scale Scientific Facility Funds of the NSFC and CAS under Contracts Nos. U1732263, U1832103, U2032111; 

%%%%%%%%%%%%%%%%%%%%%%%%%%%%%%%%%%%%%%%%%%%%%%%%%%%%%%%%%%%%%%%%%
%\bibliographystyle{apsrev4-1}
%\bibliography{./lambda_ref.bib}
%\bibliography{lambda_ref.bib}

\begin{thebibliography}{43}
\bibitem [{Zer()}]{Zernikemoment}%
  \BibitemOpen
  \href@noop {} {\ }\bibinfo {note} {\noindent The $deposited$ $energy$ is the
  total cluster energy of a shower, denoted as $E_{tot}$. The $hit$ $number$ is
  the total number of the ignited crystals of a shower. The $secondary$
  $moment$ is defined as $ \sum_{i}^{}\frac{E_{i}r_{i}^{2}}{E_{tot}}$ and the
  $Zernike$ $moment(A_{42})$ is defined as $\lvert
  \sum_{i}^{}\frac{E_{i}}{E_{tot}}f_{4,2}(\frac{r_{i}}{R_{0}})e^{im\phi_{i}}
  \rvert $ with the coefficient $f_{4,2}=4x^{4}-3x^{2}$, where $r_{i}$ and
  $\phi_{i}$ are the radial and angular seperation of crystal $i$ with respect
  to the cluster center, $E_{i}$ is the $i_{th}$ ignited crystals and $R_{0}$
  is a cutoff radius of 15~cm. The $deposition$ $shape$ is defined
  as$(E_{5\times5}-E_{3\times3})/E_{5\times5}$, where $E_{3\times3}$ and
  $E_{5\times5}$ are the deposited energies in $3\times3$ and $5\times5$
  crystals around the cluster seed, respectively.}\BibitemShut {Stop}%
\end{thebibliography}
%Control: production of eprint (-1) disabled
%
%\input{pipipsip_draft.bbl}
%%%%%%%%%%%%%%%%%%%%%%%%%%%%%%%%%%%%%%%%%%%%%%%%%%%%%%%%%%%%%%%%%
\end{document}